\documentclass[conference]{IEEEtran}
\IEEEoverridecommandlockouts
\usepackage{cite}
\usepackage{amsmath,amssymb,amsfonts}
\usepackage{algorithmic}
\usepackage{graphicx}
\usepackage{textcomp}
\usepackage{xcolor}
\def\BibTeX{{\rm B\kern-.05em{\sc i\kern-.025em b}\kern-.08em
    T\kern-.1667em\lower.7ex\hbox{E}\kern-.125emX}}

\begin{document}

\title{CNTFET quaternary multipliers are less efficient than the corresponding binary ones}

\author{\IEEEauthorblockN{ Daniel Etiemble}
\IEEEauthorblockA{\textit{Computer Science Laboratory (LISN)} \\
\textit{Paris Saclay University}\\
Orsay, France \\
de@lri.fr}
}

\maketitle


\begin{abstract}
We compare N*N quaternary digit and 2N*2N bit CNTFET multipliers in terms of Worst case delay, Chip area, Power and Power Delay Product (PDP) for N=1, N=2 and N=4. Both multipliers use Wallace reduction trees. HSpice simulations with 32-nm CNTFET parameters shows that the binary implementations are always more efficient: the 1*1 quit multiplier is far more complex than a 1*1 bit multiplier (AND gate) and generate both product and a carry terms. Even with half number of terms, the quaternary reduction tree has the same number of terms than the binary one, and uses quaternary adders that are also more complicated than the binary ones. The quaternary multipliers have larger worst case delays, more power dissipation and far more chip areas than the binary ones computing the same amount of information.
\end{abstract}


\section{Introduction}\label{sec1}
Since the 50's, many implementations of multivalued circuits have been proposed. In the last decade, most proposals used the CNTFET technology. 

Most presented implementations of ternary or quaternary circuits claim advantages of multiple valued circuits. The following quote summarizes the arguments  that may be found in most of these papers : 
``MVL circuits have potential advantages. Using MVL circuits reduces the complexity of interconnection via reducing the number
of wires since each wire carries more than one digit of data. Power consumption and area of the MVL circuits are generally less than the corresponding binary circuits due to the reduction in number of active elements \cite{Haixia}.

We examined ternary circuits in \cite{Eti1}. In this paper, we compare quaternary and binary multipliers to check the validity of the previous quote for these important combinational circuits.

The implementation of N-digit adders is easily derived from the implementation of 1-digit adders, with variants to speed-up the carry propagations. Multipliers are most complicated as they involve two steps to multiply N x N digits in the common implementations with minimal propagation delays:
\begin{itemize}
\item Multiply the ith digit by the jth digit for $0<i<N$ and $0<j<N$. It involves $N^2$ 1-digit multipliers
\item Sum the different lines of partial products. Reduction trees such as Wallace or Dada \cite{Townsend} trees are generally used. We use Wallace trees. Using other reduction techniques would not significantly change the results.
\end{itemize}
We first present the methodology that is used. Then we present the implementation a 1x1 quaternary digit multiplier. As a N*N multiplier involves both 1-digit multipliers and 1-digit adders, we successively compare  2N*2N bit multipliers with N*N quit  multipliers for N=1, N=2 and N=4. 

\section{Methodology}
\subsection{Why CNTFET technology?}

CNTFET technology uses field-effect transistors that have a single carbon nanotube or an array of carbon nanotubes as the channel material instead of bulk silicon in the traditional MOSFETs. The MOSFET-like CNTFETs having p and n types look the most promising ones. The technology has advantages and drawbacks:
\begin{itemize}
\item CNTFETs have variable threshold voltages (according to the inverse function of the diameter). This is a big advantage compared to CMOS for which different masks are needed to get different threshold voltages. 
\item Among advantages, high electron mobility, high current density, high tranductance can be quoted.
\item Lifetime issues, reliability issues, difficulties in mass production and production costs are quoted as disadvantages.
\item CNTFET technology is far from being a mature one. In 2019, a 16-bit RISC microprocessor has been built with 14,000 CNFET transistors \cite{Hills}. While this is an advance for CNTFET technology, we may observe that the Intel 8086 CPU, which was a 16-bit microprocessor, has been launched in 1978 with 29,000 transistors, more than 40 years ago!
\end{itemize} 
However, as CMOS circuits and CNTFET ones have basically the same circuit styles, CNTFETs can be used to propose a new implementation of quaternary operators and compare it with previous published proposals such as \cite{Moaiyeri} and \cite{Rahmati}.

The 32nm CNTFET parameters of Stanford library \cite{17} are used for HSpice simulations. These simulations are done at 25°C temperature. As both binary and quaternary simulations use the same technological parameters, there is no reasons for getting different qualitative comparison results with different temperature values.

\subsection{Propagation delays}
Generally, propagation delays are presented as an average of the delays corresponding to all combinations of input transitions. This presentation could be confusing. For the combinational multipliers that we consider, the multiplicand and multiplicator data are simultaneously available. The significant propagation delays are the worst case delay from inputs to the final output. 
We will only present these critical delays.
\subsection{Power and Energy dissipation}
 Power and pdp (Power Delay Product) directly depends on the duration of the input signals. It is important to use the same input signals for all designs.

\subsection{Chip area}
Without drawing the layout of the circuits, there is no technique to evaluate the chip area. 
We use a rough evaluation of the chip area by summing the diameters of all the used transistors by each circuit. This rough evaluation is a little bit better than the transistor count. In this paper, we use the diameter values presented in Table \ref{Diameter}.

\begin{table}
\centering
\caption{Transistor diameters}
\begin{tabular}{|c|c|c|c|c|c|c|c|c|}
  \hline
n&Diameter (nm)&$|Vth|$ (V)\\
  \hline
  8&0.626&0.696\\
10&0.783&0.557\\
13&1.018&0.428\\
19&1.487&0.293\\
29&2.27&0.192\\
37&2.896&0.150\\
  \hline
\end{tabular}
\label {Diameter}
\end{table}

\section{Comparing a quaternary 1-digit multiplier with a 2-bit binary one}

\subsection{Quaternary 1-digit multiplier}
\label{L1}
Table \ref{T1} shows the truth table of a 1-digit quaternary multiplier.
From Table \ref{T1}, we can observe that:
\begin{itemize}
\item When A = 0 then QM = 0 and QC = 0
\item When A = 1 then QM = B and QC = 0/0/0/1 for B = 0/1/2/3
\item When A = 2 then QM = 0/2/0/2 and QC = 0/0/1/1 for B = 0/1/2/3
\item When A = 3 then QM = 0/3/2/1 and QC= 0/0/1/2 for B = 0/1/2/3
\end{itemize}
\begin{table}
\centering
\caption{Truth table of a quaternary multiplier}
\begin{tabular}{|c|c||c|c|c|c|c||c|c||c|c|}
  \hline
A&Bi&QS&QC&&A&Bi&QM&QC\\
\hline
0&0&0&0&&2&0&0&0\\
0&1&0&0&&2&1&2&0\\
0&2&0&0&&2&2&0&1\\
0&3&0&0&&2&3&2&1\\
1&0&1&0&&3&0&0&0\\
1&1&2&0&&3&1&3&0\\
1&2&3&0&&3&2&2&1\\
1&3&0&1&&3&3&1&2\\
  \hline
\end{tabular}
\label {T1}
\end{table}

\begin{figure}[htbp]
\centerline{\includegraphics  [width =3 cm]{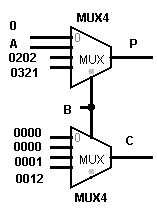}}
\caption{1-digit quaternary multiplier}
\label{QM1}
\end{figure}

\begin{figure}[htbp]
\centerline{\includegraphics  [width =6 cm]{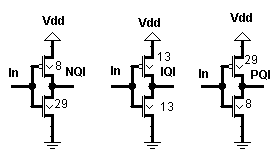}}
\caption{Threshold detectors}
\label{QDEC4}
\end{figure}

Using the same mux-based technique as in \cite{Roosta}, the Product and Carry circuits are shown in Fig. \ref{QM1}. The quaternary input values are decoded into binary values by NQI, IQI and PQI functions according to Table \ref{Q2B} in which binary values are 0 and 3. NQI, IQI and PQI outputs are provided by 3 inverters having 3 different threshold levels (Fig. \ref{QDEC4}).

\begin{figure}[htbp]
\centerline{\includegraphics  [width =4 cm]{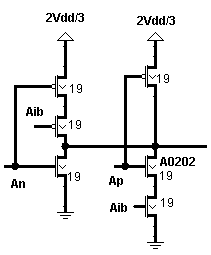}}
\caption{0202 Operator}
\label{0202}
\end{figure}

\begin{figure}[htbp]
\centerline{\includegraphics  [width =4 cm]{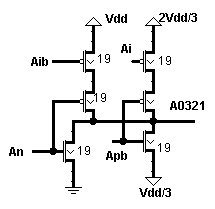}}
\caption{0321 Operator}
\label{0321}
\end{figure}

\begin{figure}[htbp]
\centerline{\includegraphics  [width= 4cm]{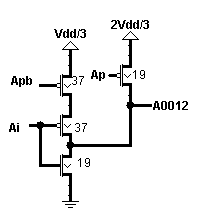}}
\caption{0012 Operator}
\label{0012}
\end{figure}

\begin{figure}[htbp]
\centerline{\includegraphics[width=6cm]{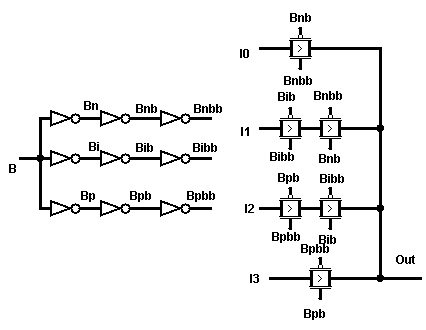}}
\caption{4-input MUX with quaternary control}
\label{QMUXDE}
\end{figure}

Operators 0202 and 0321 are respectively shown in Fig. \ref{0202} and \ref{0321}.
Operators 0001 and 0012 are derived from the circuits implementing the  NQI, IQI and PQI functions.  A0012 circuit is shown in Fig. \ref{0012}. Circuits for A0001/A0011/A0111 are inverters with $V_{dd}$/3 power supply with Ap/Ai/An inputs. The 4-input MUX with quaternary control is shown in Fig. \ref{QMUXDE}. It should be noticed that the threshold detectors shown in Fig. \ref{QDEC4} have low driving capability. This is why Bn, Bi and Bp outputs in Fig. \ref{QMUXDE} are buffered by two inverters to get Bnbb, Bibb and Bpbb with sufficient fan-out.

\begin{table}
\centering
\caption{Truth table of decoder circuits}
\begin{tabular}{|c||c|c|c|}
  \hline
 IN&NQI&IQI&PQI\\
\hline
 0&3&3&3\\
 1&0&3&3\\
 2&0&0&3\\
 3&0&0&0\\
  \hline
\end{tabular}
\label {Q2B}
\end{table}

\subsection{2-bit binary multiplier} \label{HA}
\begin{figure}[htbp]
\centerline{\includegraphics  [width =8 cm]{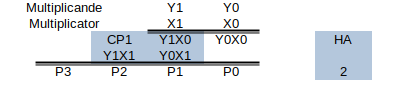}}
\caption{2*2-bit multiplication}
\label{2X2mul}
\end{figure}

\begin{figure}[htbp]
\centerline{\includegraphics  [width =6 cm]{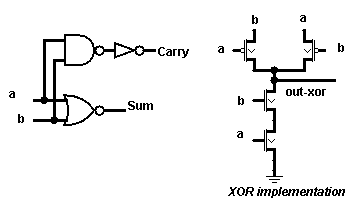}}
\caption{Binary Half Adder}
\label{BHA}
\end{figure}

The basic scheme of a 2*2 bit multiplication is shown in Fig. \ref{2X2mul}. The corresponding multiplier has 4 AND gates to compute the YjXi products and two half adders (HA) to sum the partial products. Fig. \ref{BHA} presents the binary half adder.

\subsection{Performance of the 1-quit multiplier and the 2-bit one}
These two multipliers compute the same amount of information. While the 1-quit multiplier has a 0.9V $V_{dd}$ power supply, the 2-bit binary multipliers can use different power supplies: we use 0.9V and 0.45V power supplies. This last one can roughly reduce by x4 the dynamic power dissipation. 

\begin{figure}[htbp]
\centerline{\includegraphics  [width =10 cm]{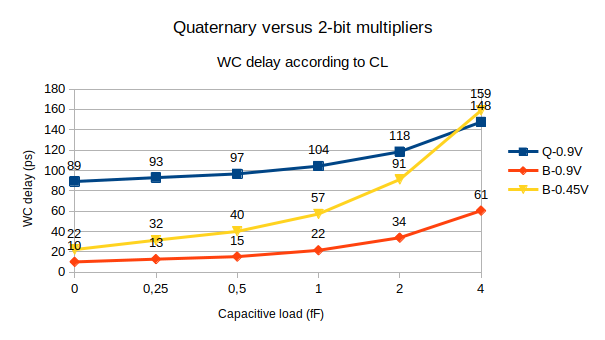}}
\caption{WC delays for 1-quit and 2-bit multipliers}
\label{1T2BDelay}
\end{figure}

Fig. \ref{1T2BDelay} presents the worst case delays according to capacitive load for the quaternary multipliers and the two binary ones. The quaternary multiplier is outperformed by the 0.9V binary one. The difference ranges from x8.9 to x2.5 when the capative load ranges from 0 to 4fF. Except for CL=4fF, the WC delay is also greater for the quaternary multiplier than for the 0.45V binary one.

\begin{figure}[htbp]
\centerline{\includegraphics  [width =10 cm]{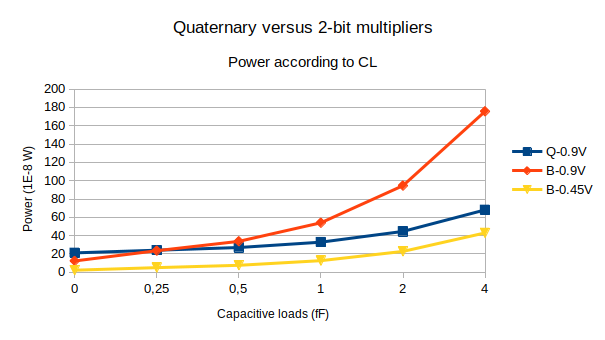}}
\caption{Power dissipation for 1-quit and 2-bit multipliers}
\label{1T2BPower}
\end{figure}

\begin{figure}[htbp]
\centerline{\includegraphics  [width =10 cm]{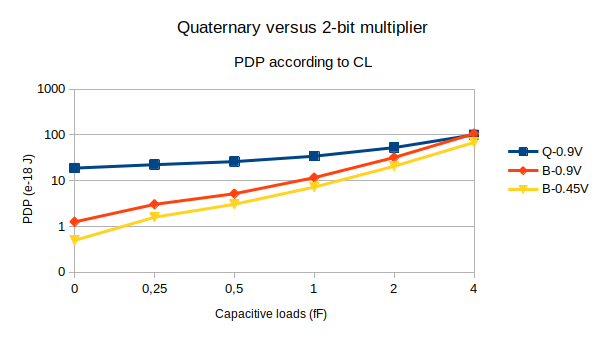}}
\caption{PDP for 1-quit and 2-bit multipliers}
\label{1T2BPDP}
\end{figure}

Fig. \ref{1T2BPower} presents the power dissipation according to capacitive load. The quaternary multiplier dissipate less than the 0.9V binary one, but dissipate more than the 0.45V binary one. Fig. \ref{1T2BPDP} presents the PDP according to capacitive load. The binary multipliers are more sensitive to capacitive load than the quaternary one, but PDP is lower for the binary ones except for CL=4fF.

$\Sigma{Di}=71$ for the 2*2 bit multiplier and $\Sigma{Di}=132$ for the quaternary one.

\section{Comparing a 2*2 quit multiplier with a 4*4 bit one}
\subsection{A 2*2 quit multiplier}
\begin{figure}[htbp]
\centerline{\includegraphics  [width = 9 cm]{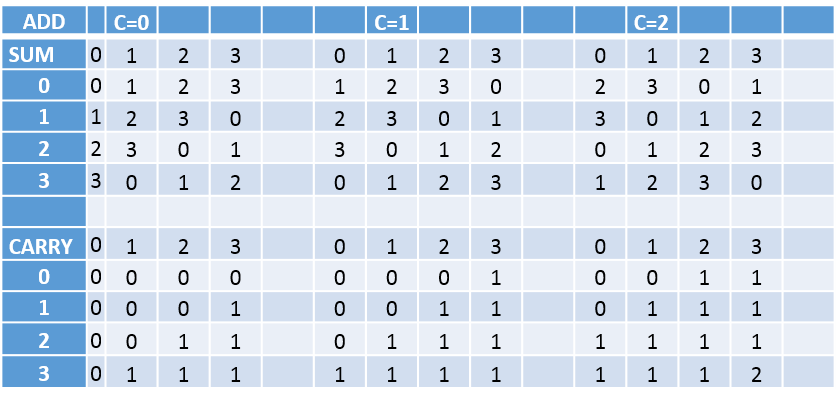}}
\caption{Truth Table of Quaternary Adders.}
\label{TTQADD}
\end{figure}

\begin{figure}[htbp]
\centerline{\includegraphics[width=8cm]{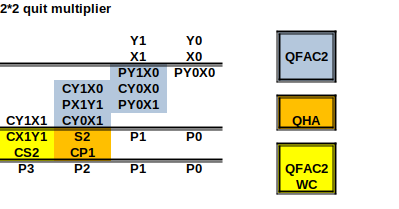}}
\caption{Multiplying 2*2 quits}
\label{2*2qmul}
\end{figure}

\begin{figure}[htbp]
\centerline{\includegraphics[width=8cm]{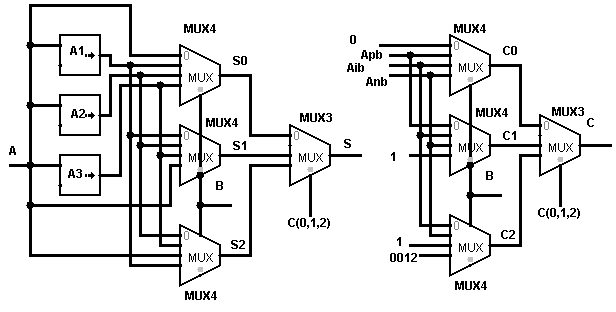}}
\caption{Quaternary adder with ternary carry values}
\label{QFAC2}
\end{figure}

\begin{figure}[htbp]
\centerline{\includegraphics[width=8cm]{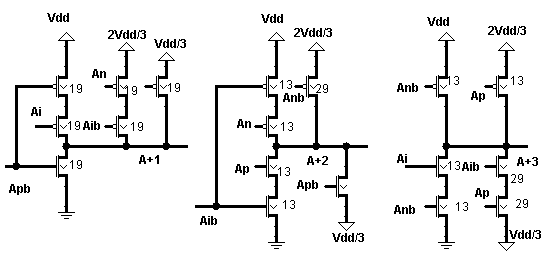}}
\caption{A¹, A² and A³ circuits}
\label{A1A2A3}
\end{figure}

The basic scheme of a 2*2 quit multiplication is shown in Fig. \ref{2*2qmul}. The corresponding multiplier has four 1-quit multipliers that generate four product terms Pyixj and four carry terms Cyixj. There is one stage of reduction tree and a final add. It should be noticed that the carry terms have ternary value (0,1,2) according to Table \ref{T1}. It means that a special adder (QFAC2) is needed with ternary carry values. The truth table of QFAC2 is shown in Fig. \ref{TTQADD}. The corresponding circuit is shown in Fig. \ref{QFAC2}. Operators A1, A2 and A3 are shown in Fig. \ref{A1A2A3}. Operator 0012 has been presented in Fig. \ref{0012}. Apb, Aib and Anb are the binary complements of Ap, Ai and An. QHA is the quaternary half adder. QFAC2WC is a quaternary adder with ternary carry input for which no carry output is computed as no carry is generated by the left adder of the final add. 

\subsection{A 4*4 bit multiplier}
\begin{figure}[htbp]
\centerline{\includegraphics[width=10cm]{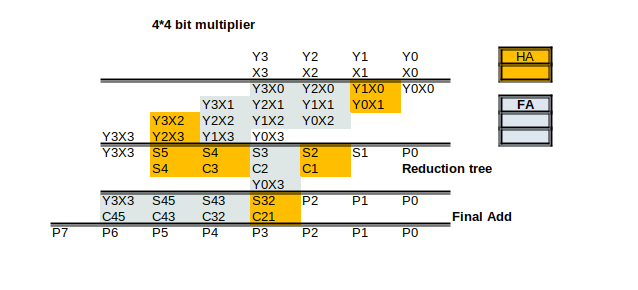}}
\caption{Multiplying 4*4 bits}
\label{4*4bmul}
\end{figure}

\begin{figure}[htbp]
\centerline{\includegraphics[width=6cm]{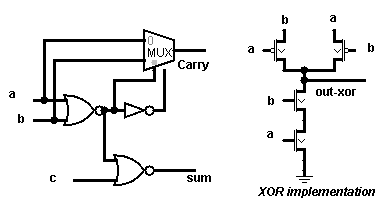}}
\caption{14T Binary Full Adder}
\label{14T}
\end{figure}

The basic scheme of a 4*4 bit multiplication is shown in Fig. \ref{4*4bmul}. The corresponding multiplier has 16 AND gates to compute XiYj, two stages of Wallace reduction trees and a final add. It uses 6 binary HAs and 7 binary FAs. The HA scheme has been presented in \ref{HA}. The FA scheme is presented in Fig. \ref{14T}.

\subsection{Performance of the 2-quit multiplier and the 4-bit one}
\begin{figure}[htbp]
\centerline{\includegraphics[width=9cm]{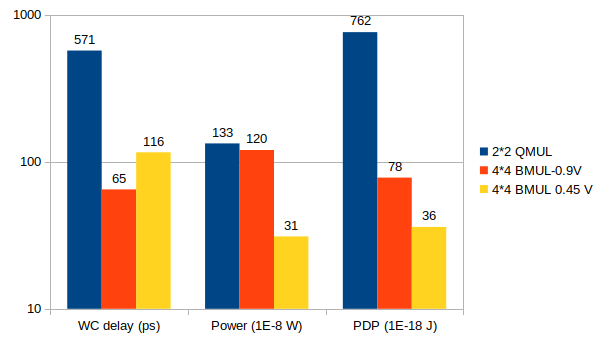}}
\caption{Comparing a 2*2 quit multiplier with a 4*4 bit one}
\label{C2Q4B}
\end{figure}

\begin{figure}[htbp]

\centerline{\includegraphics[width=9cm]{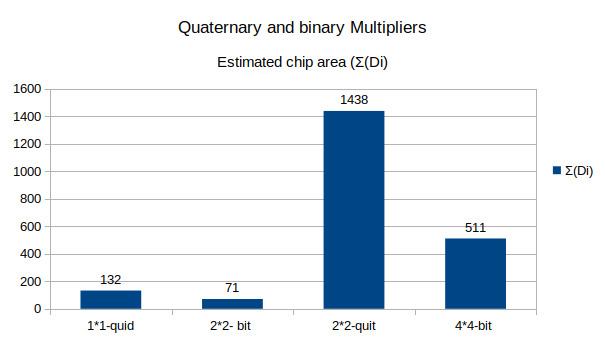}}
\caption{Estimated chip area for 2*2 quit multiplier and a 4*4 bit one}
\label{C2Q4Barea}
\end{figure}

For any Wallace multiplier, the worst case delay corresponds to the max delay through the longest path through the successive stages of the reduction tree plus the subsequent propagation delay through the final adder.
The corresponding WC delay is exhibited by the following input values:
\begin{itemize}
\item In the binary case, 1111*1111=111000001 and 1111*1101=10100101. Multiplying 1111*11x1 and switching x ($0\rightarrow 1 \rightarrow 0$) leads to the WC propagation delay between x and P6 output.
\item In the quaternary case, 23*33=2211 and 23*23=1321. Multiplying 23*x3 and switching x ($2\rightarrow 3 \rightarrow 2$) leads to the WC propagation delay between x and P3 output. 
\end{itemize}
Due to the large simulation times, the results are only provided with a 2fF capacitive load on every output. Fig. \ref{C2Q4B} presents the WC delays, the Power dissipation and the PDP for the quaternary multiplier and the two versions of the binary multipliers.
It turns out that the WC delay of the quaternary adder is respectively x8.8 and x4.9 the WC delay of the binary versions. While power dissipation is close for the 0.9V power supplies, the quaternary power is roughly x4 the power dissipation of the 0.45V binary version. The PDP ratios are x10 and x22 greater for the quaternary multiplier compared to the two versions of the binary ones. 
These results are not surprising:
\begin{itemize}
\item The 1-quit multiplier is far more complicated than a AND gate. There are 4 times less 1-quit multipliers than AND gates, but the 1-quit multipliers generate both  product and carry terms, which means that there is the same number of lines of terms even with quaternary lines that have two times less terms.
\item The quaternary adders used in the reduction tree and the final add are far more complicated than binary half adders and full adders.
\item As shown in Fig. \ref{C2Q4Barea}, the estimated chip area is x2.8 larger for the  2*2 quit multiplier than for the 4*4 bit one, while the ratio is x1.9 for the 1*1 quit multiplier compared to the 2*2 bit one.

\end{itemize}



\section{Comparing a 4*4quit multiplier with a 8*8 bit one}

\subsection{The Wallace reduction trees}

\begin{figure}[htbp]
\centerline{\includegraphics  [width = 9 cm]{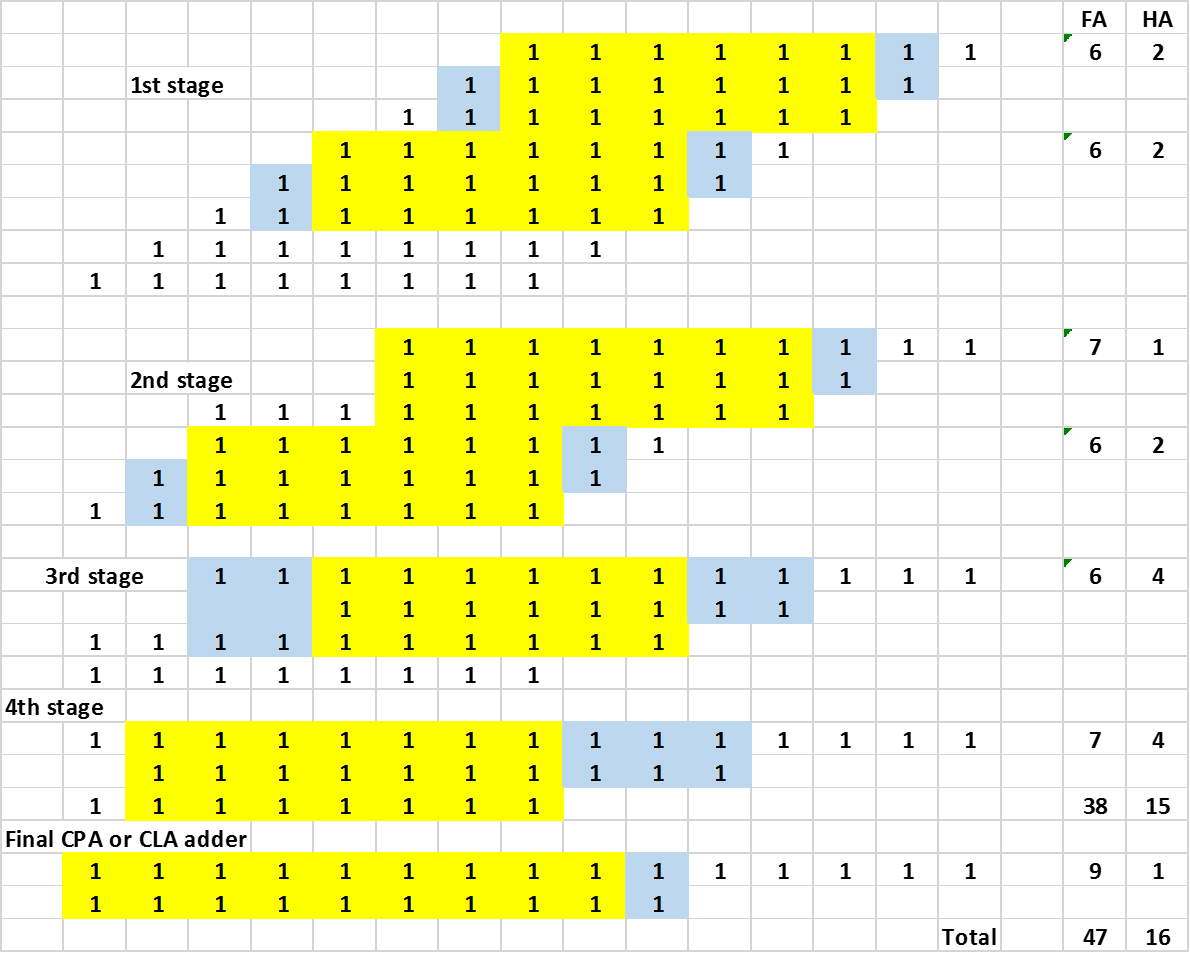}}
\caption{8*8 Wallace Multiplier} 
\label{W88}
\end{figure}

Fig. \ref{W88} presents the 8*8 bit Wallace multiplier. There are 64 AND gates (1*1 bit multiplier), 38 1-bit adders (FAs) and 15 1-bit half adders (HAs) for the reduction tree. A final 10-bit Carry Propagate Adder (CPA) uses 9 FAs and 1 HA.
The critical path determining the WC delay consists in a series of 2 FAs and 2 HAs in the four stage of reduction tree plus the propagation delay in the final adders (1 HA and 9 FAs). 
64 AND gates, 47 FAs and 16 HAs are used. $\Sigma{Di}=2377$.

\begin{figure}[htbp]
\centerline{\includegraphics  [width =8 cm]{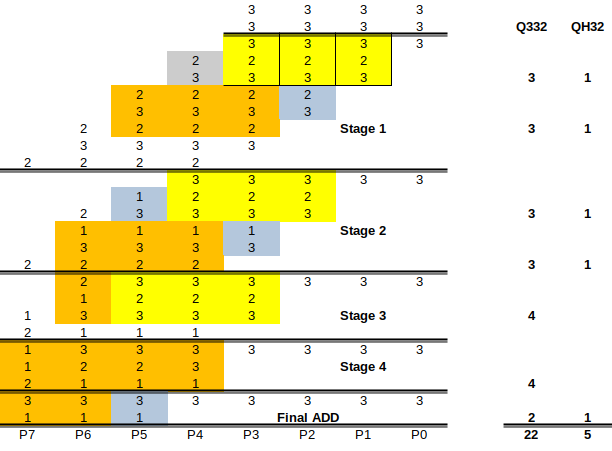}}
\caption{4x4 Wallace Tree}
\label{WQ44}
\end{figure}

Fig. \ref{WQ44} presents the 4*4 quit Wallace multipiers. There are 16 1*1 quit multipliers that generates a product and a carry terms. So, there are 8 lines of terms to reduce, exactly as for the 8*8 bit multipliers. Both multipliers have 4 stages of reduction tree. In Fig. \ref{WQ44}, 3 corresponds to quaternary values (max=3), 2 to ternary values (max=2) and 1 to binary values (max=1). Two types of adders are used:
\begin{itemize}
\item Q332 has quaternary inputs, ternary carry inputs, quaternary sum output and ternary carry output. It corresponds to QFAC2 (Fig. \ref{QFAC2}). Carry outputs are either ternary or binary according to the corresponding inputs.
\item QHA32 is the quaternary half adder.
\item Layout may be complicated as quaternary inputs must be distinguished from the ternary carry inputs.
\end{itemize}.

The 4*4 quit multiplier uses 16 1*1 quit multipliers, 22 QFAC2s and 5 QHAC2. $\Sigma{Di}=7530$
The 4*4 quit multiplier has x3.2 more estimated chip area than the 8*8 bit one. The ratio was x2.8 for 2*2 quit and 4*4 bit multipliers. 

\subsection{WC propagation delays}
For these multipliers, HSPICE simulations of the completed 4*4 quit and 8*8 bit multipliers would need several hours. We evaluate the WC propagation delays by simulating only the WC path in the Wallace tree.

For the binary tree, the WC path corresponds to three sum propagation delays and 1 carry delay vertically in the reduction tree, followed by 10 carry propagation delays horizontally in the final add. The corresponding propagation delay (without the AND delay) is 312 ps ($V_{dd}$=0.9V) and 799 ps ($V_{dd}$=0.45V) when P14 and P15 have a 2fF capacitive load.

For the quaternary tree, the critical path  consists in a series of 4 QFAC2 (one per stage) plus the propagation in the final ADD (1 QHA and two QFCA2). The propagation delay (without the 1-quit multiplier delay) is 646 ps when P7 and P6 have a 2fF capacitive load. The 1-trit multiplier delay (CL=2fF) is 118 ps, which is far more than a AND gate propagation delay.
Except for Power when compared to binary multiplier with $V_{dd}$=0.9V and for WC delay compared to the binary one with $V_{dd}$=0.45V, the quaternary 4*4 quit multiplier is less efficient than the binary ones. It is outperformed by the 0.45V binary version for PDP, and by both binary adders for chip area.
We precise that these figures are less precise than the previous one as only the WC path (and not the complete multipliers) have been simulated.

\begin{figure}[htbp]
\centerline{\includegraphics  [width =8 cm]{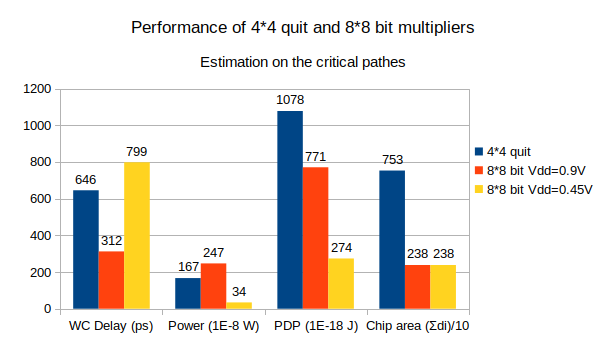}}
\caption{Performance of 4*4 quit and 8*8 bit Wallace tree}
\label{C4488}
\end{figure}

\section{Discussing results}
\begin{figure}[htbp]
\centerline{\includegraphics  [width =8 cm]{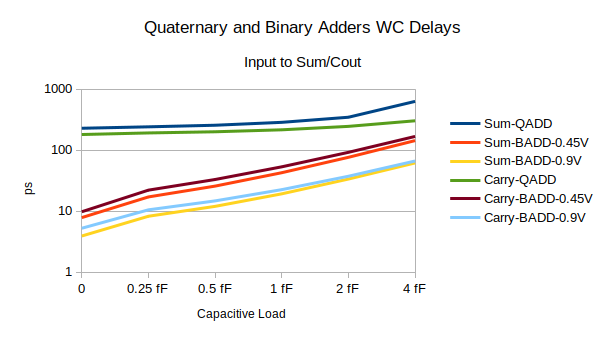}}
\caption{WC input to sum/output delays of binary and quaternary adders according to capacitive loads}
\label{QBCWID}
\end{figure}

\begin{figure}[htbp]
\centerline{\includegraphics  [width =8 cm]{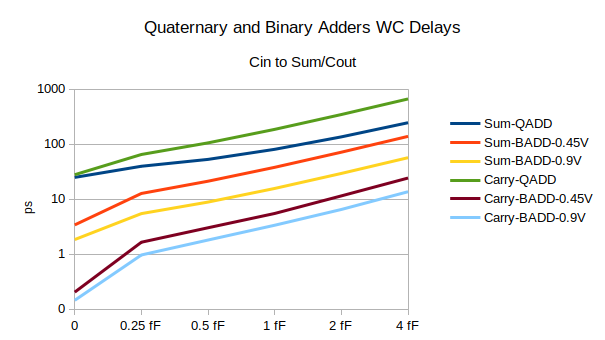}}
\caption{WC Cin to sum/Cout delays of binary and quaternary adders according to capacitive loads}
\label{QBCWCD}
\end{figure}

The first drawback of the quaternary multipliers is the 1*1 quit multiplier that generates both product and carry terms. It is far more complicated than a binary AND gate.
\begin{itemize}
\item $\Sigma{Di}=8.9$ for a AND gate and $\Sigma{Di}=132$ for a 1*1 quit multiplier. This x15 ratio (chip area) prohibits from using the fact that there are 4 times less 1*1 quit multipliers than AND gates. The WC delays for the 1*1 quit multplier were presented in Fig. \ref{1T2BDelay}. They are far greater than the corresponding AND gate delays.
\item The number of lines to be reduced is the same for both multipliers, even if there are two times less terms per line in the quaternary case. Less adders are used in the quaternary case, but the quaternary adders are far more complicated, which means larger chip area and larger propagation delays. 
\end{itemize} 
$\Sigma{Di}=18$ for the binary HA  and $\Sigma{Di}=32$ for the binary FA. $\Sigma{Di}=83$ for the quaternary HA  and $\Sigma{Di}=227$ for the quaternary FA. The ratios are 
x4.6 for HAs and 7.1 for FAs. For the 4*4 quit and 8*8 bit multipliers, the chip area ratios of HAs and FAs between the quaternary and binary ones are respectively 5/16=0.31 and 22/47=0.47. The smaller number of quaternary adders cannot compensate the greater chip area ratios.
The Input to Sum/Cout WC delays of the quaternary adder with ternary carry and the binary adder are presented in Fig. \ref{QBCWID}. While the binary adders are more sensitive to capacitive loads, the delays are always significantly lower. The corresponding Cin to Sum/cout WC delays are presented in Fig. \ref{QBCWCD}. Again, the binary delays are significantly lower: the difference is one order of magnitude for the 0.45 V power supply of the binary adder.

\section{Concluding remarks}
The most efficient implementation of quaternary 1-digit adders and multipliers uses the multiplexer approach. Even with this approach, these adders and multipliers are more complicated than the corresponding binary ones in terms of worst case delays, estimated chip area, power and PDP. The 1*1 quit multiplier generates both product and carry terms, which leads to the same number of lines in the Wallace reduction tree in N*N quit and 2N*2N bit multipliers computing the same amount of information. More, with quaternary circuits using $V_{dd}$ power supply, some parts of the circuits operate with $V_{dd}$/3 power supplies. This means that the corresponding binary circuits can operate with $V_{dd}$/3 power supplies, which means less power dissipation. We used $V_{dd}$/2 power supply to illustrate this point.
The quaternary multipliers have always greater WC delays, Power, PDP than the corresponding binary ones, except for Power with the 1*1 quit multiplier. However, such a multiplier is never used alone in a design. Power supplies smaller than $V_{dd}$ for binary multipliers significantly decrease power and PDP.
The supposed advantages of multivalued circuits quoted in section \ref{sec1} are thus disproved.

\end{document}